\documentclass{svjour3}     
\smartqed  
\usepackage{graphicx}
\usepackage{colortbl}
\usepackage{xcolor}
\usepackage{ulem}
\usepackage{wasysym}
\usepackage{amssymb, latexsym, textcomp,  textcomp}
\journalname{JLTP}
\begin{document}

\title{Cryogenic resonant amplifier for electron-on-helium image charge readout}

\titlerunning{Cryogenic resonant amplifier for electron-on-helium image charge readout}        

\author {Mikhail Belianchikov$^1$$^\dagger$, Jakob A. Kraus$^2$, and Denis Konstantinov$^1$}


\institute{1. Quantum Dynamics Unit, Okinawa Institute of Science and Technology (OIST) Graduate University, Tancha 1919-1, Okinawa 904-0412, Japan.\\ 
}

\institute{1. Okinawa Institute of Science and Technology, Tancha 1919-1, Okinawa 904-0412, Japan \\
2. Fachbereich Physik, Universit\"at Konstanz, DE-78457 Konstanz, Germany \\ 
 \email{m.belianchikov@oist.jp} }


\maketitle
\begin{abstract}
An electron-on-helium qubit is a promising physical platform for quantum information technologies. Among all the “blueprints” for the qubit realization, a hybrid Rydberg-spin qubit seems to be a promising one towards quantum computing using electron spins. The main technological challenge on the way to such qubits is a detection of fA range image current induced by Rydberg transition of a single electron. To address this problem we aim to use a tank LC-circuit in conjunction with a high impedance and low power dissipation cryogenic amplifier. Here, we report our progress towards realization of a resonant image current detector with a home-made cryogenic amplifier based on FHX13LG HEMT. We present a detailed characterization of the transistor at room and cryogenic temperatures, as well as details of the amplifier design and performance. At the power dissipation level of amplifier well below 100~${\rm \mu}$W the measured voltage and current noise level is 0.6~nV/$\sqrt{\rm Hz}$ and below 1.5~fA/$\sqrt{\rm Hz}$, respectively. Based on the actual image current measurements of the Rydberg transition in a many-electron system on liquid helium, we estimate SNR=8 with the measurement bandwidth 1 Hz for the detection of a single-electron transition, providing the noise level at the output is solely determined by the noise of the amplifier.              

\end{abstract}

\section{Introduction}
\label{intro}  
The current stage of development of the quantum technologies generates a plethora of pioneering research towards finding the best physical system for the quantum computer realization. Such well-established experimental platforms as superconducting circuits, electron spins in semiconductors, and trapped ions already demonstrate high-fidelity multi-qubit operations, but suffer from poor scalability to a larger number of qubits necessary for implementation of the error-correction codes~\cite{JonePRX2012}.  Towards the latter, there is a revival of significant interest in the system of surface-state electrons on liquid helium and other cryogenic substrates~\cite{Monarkha}. Offering an extremely clean environment and a mobile qubit platform, electrons on helium holds a promise to be a good candidate for scalable qubits~\cite{DykmPRB2003,LyonPRA2004,SchuPRL2010,JinQST2020}. One of the main challenges towards building such qubits is the detection of the quantum state of a single electron. Among the proposed solutions are destructive ionization~\cite{PlatSci1999}, single-electron transistor (SET)~\cite{DevoNat2000,PapaAPL2005,RousPRB2009}, and dispersive readout using a superconducting coplanar waveguide resonator~\cite{YangPRX2016}. Recently, the latter method has been successfully realized to detect the quantized states of the in-plane motion for electrons on superfluid helium and solid neon~\cite{KoolNat2019,ZhouNat2022}. There is also a significant interest to the detection of the quantized states of the electron's vertical motion, the so called Rydberg states~\cite{LeaPRL2002}. Most recently, the artificially introduced coupling of such states to the electron spin was proposed to facilitate the spin-based quantum computing with electrons on liquid helium~\cite{Kawa2023}. In particular, it was shown that the spin-state readout of such a hybrid Rydberg-spin qubit can be accomplished by detecting the transition between the Rydberg states. The typical transition frequency for the Rydberg states of electrons on helium is above 100~GHz, therefore the method of the dispersive readout with a coplanar waveguide resonator is not applicable. For the purpose of detecting the Rydberg transition of a single electron a new method of the image charge detection was proposed and successfully realized in a many-electron system on liquid helium~\cite{KawaPRL2019}. Subsequently, a route towards realization of the Rydberg transition detection of a single electron was demonstrated by employing electrons on superfluid helium confined in a microchannel structure~\cite{ZouNJP2022}.  

The image charge readout could be considered as the most conceptually and technologically simple and, hence, feasible. It is based on the detection of the image current induced in a conducting electrode by the motion, either classical or quantized, of a charged particle located in the electrode proximity. In particular, this approach is widely used for the detection of a single electron or an ion trapped in the Penning trap~\cite{VanDPRL1977,WineJAP1975}. At the heart of the method used for the trapped particle detection is a parallel LC tank circuit formed by a helical resonator and the stray capacitance of the experimental setup. Usually, superconducting materials are used for the helical resonator to achieve the highest possible quality factor $Q$ of the tank circuit. At resonance, such a circuit provides a high input impedance $\propto Q$ for the measured image current to accomplish the trans-impedance conversion. In conjunction with a high input-impedance voltage amplifier, such circuit forms a resonant amplifier which is capable of measuring an RF current signal in fA range at the cost of the measurement bandwidth~\cite{JeffRSI1993}.
 
Here, we aim to adopt this resonant amplification scheme for the readout of the image charge signal from the Rydberg transitions of electrons on liquid helium. Based on the literature review, we have chosen the FHX13LG High-Electron-Mobility-Transistor (HEMT)~\cite{HEMT} to build a high input-impedance cryogenic amplifier. By combining it with a fabricated  helical resonator we designed a cryogenic resonant amplifier capable to detect the Rydberg transition in a many-electron system on liquid helium. By analyzing the measured signal and comparing it with the measured noise characteristics of the amplifier we conclude that the detection of a single-electron Rydberg transitions using such a setup is feasible. Below, we present the details of the amplifier design, its characterization, and the Rydberg-transition detection.  

\section{Methods}
\label{methods}

\subsection{HEMT characterization}
\label{sec:hemt}

\begin{figure}
	\includegraphics[width=1.0\textwidth]{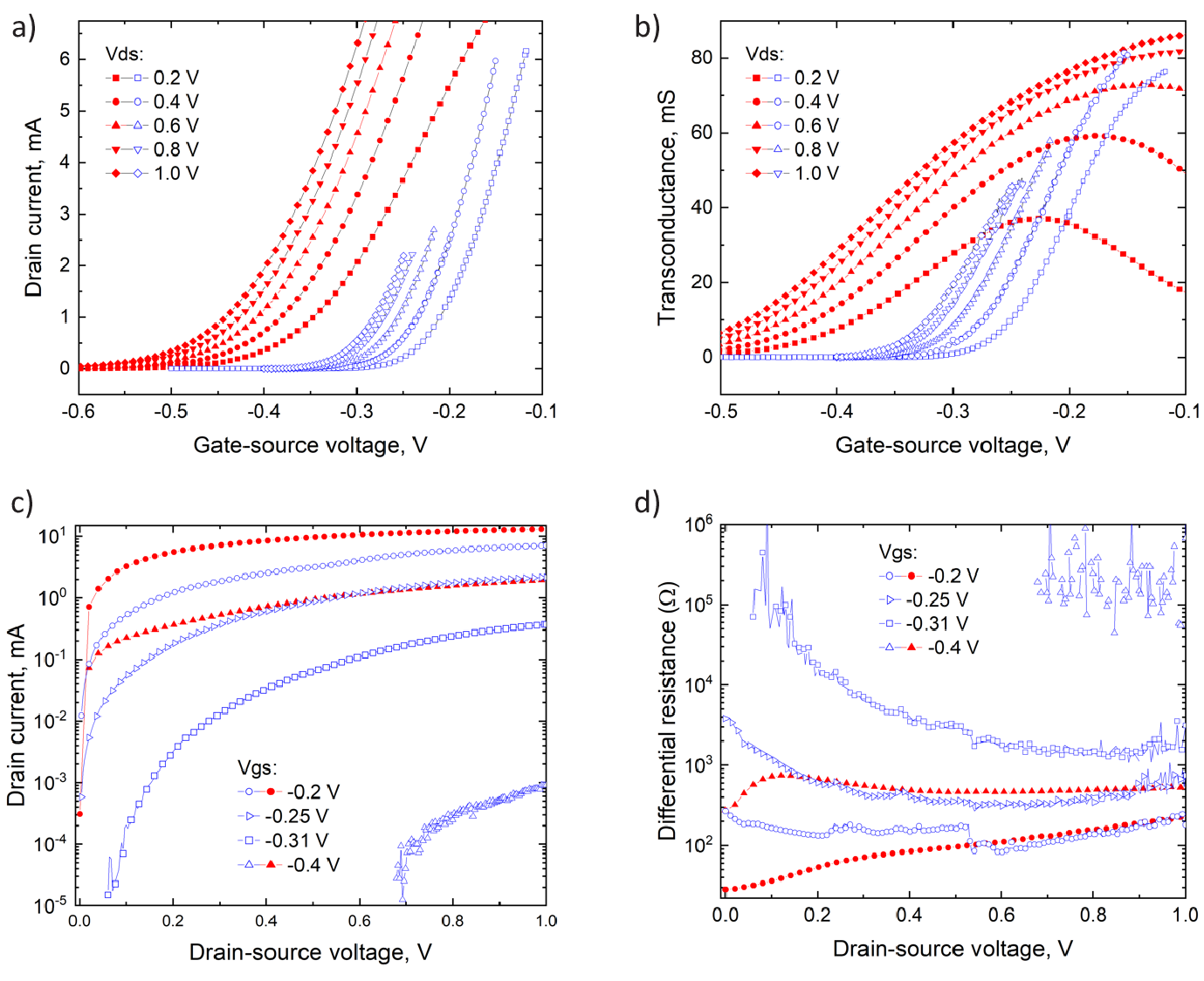}
	\caption{(color online) (a) Drain current $I_d$ versus gate-source voltage $V_{gs}$, (b) transconductance $(\partial I_d/\partial V_{gs})$ versus gate-source voltage, (c) drain current $I_d$ versus drain-source voltage $V_{ds}$, and (d) output differential resistance $(\partial I_d/\partial V_{ds})^{-1}$ versus drain-source voltage of FHX13LG HEMT measured at $T=300$ (red closed symbols) and 0.15~K (blue opened symbols).}
	\label{fig:1}	 
\end{figure}

The FHX13LG HEMT has been previously employed for making a cryogenic trans-impedance amplifier used for the image charge detection of the axial motion of a single electron in the Penning trap~\cite{UrsoPRL2005}. However, no published data about its cryogenic performance can be found. Considering that the performance of HEMT transistor may significantly vary with the temperature~\cite{PospIEEE1988}, we have measured the DC characteristics of FHX13LG HEMT at different temperatures. The cryogenic measurements were done by mounting the transistor on PCB placed at the mixing chamber plate of a dilution refrigerator. To improve thermalization of the transistor during measurements, the grounded source leads of the transistor were connected to the mixing chamber plate with a copper heat sinks. Fig.~\ref{fig:1}(a) shows the drain current $I_d$ versus the gate-source voltage $V_{gs}$ measured at $T=300$ (solid symbols) and 0.15~K (open symbols) for different values of the drain-source voltage $V_{ds}$. It was found that the measured characteristics do not change significantly at temperatures below 4~K. Based on the measured $IV$-curves, the transconductance $g_m=(\partial I_d/\partial V_{gs})$ is determined and plotted in Fig.~\ref{fig:1}(b).   

\begin{figure}
	\includegraphics[width=1\textwidth]{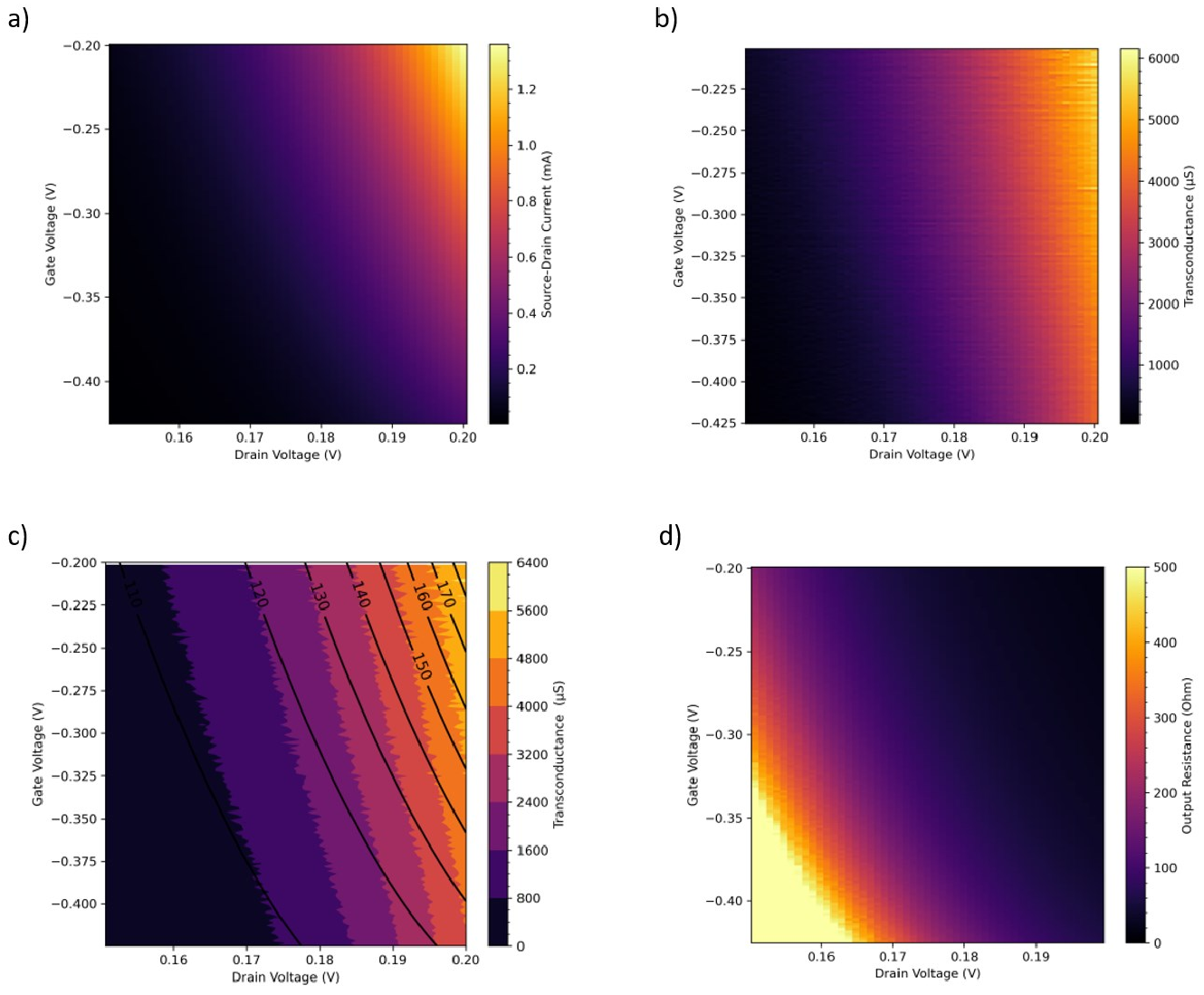}
	\caption{(color online) Color maps of (a) drain current $I_d$, (b and c) transconductance $g_m$, and (d) output differential resistance $r_0$ versus the drain and gate voltages. Panel (c) shows the contour lines of the measured temperature overlapped with the transconductance data.}
	\label{fig:2}	 
\end{figure}

For the proper impedance matching of the transistor to the readout resonant circuit the small-signal output resistance of the transistor should be known. For this purpose we measured the drain current $I_d$ versus drain-source voltage $V_{ds}$ at different values of the gate-source voltage $V_{gs}$. The results are shown in Fig.~\ref{fig:1}(c). The small-signal output (differential) resistance is then determined as $r_o=(\partial I_d/\partial V_{ds})^{-1}$ and is plotted in Fig.~\ref{fig:1}(d). Since the measured $r_o$ varies significantly with DC voltages, the impedance matching of the amplifier output should be done for a chosen transistor DC biasing.  

Figs.~\ref{fig:2}(a-d) show the color maps of $I_d$, $g_m$ and $r_o$ versus the drain-source voltage $V_{ds}$ and gate-source voltage $V_{gs}$ measured at the temperature of the mixing chamber plate in the range from 100 to 200~mK. In this measurements, the temperature of the plate was determined by the power dissipation of the transistor $I_dV_{ds}$. Fig.~\ref{fig:2}(c) also shows the mixing chamber plate temperature (contour lines) overlapped with the transconductance data. From this plot we can clearly identify that at a fixed temperature of the mixing chamber plate the higher transconductance values can be achieved for lower (more negative) gate voltages and higher drain voltages. 

\subsection{Resonant amplifier design}

\begin{figure}
	\includegraphics[width=1.0\textwidth]{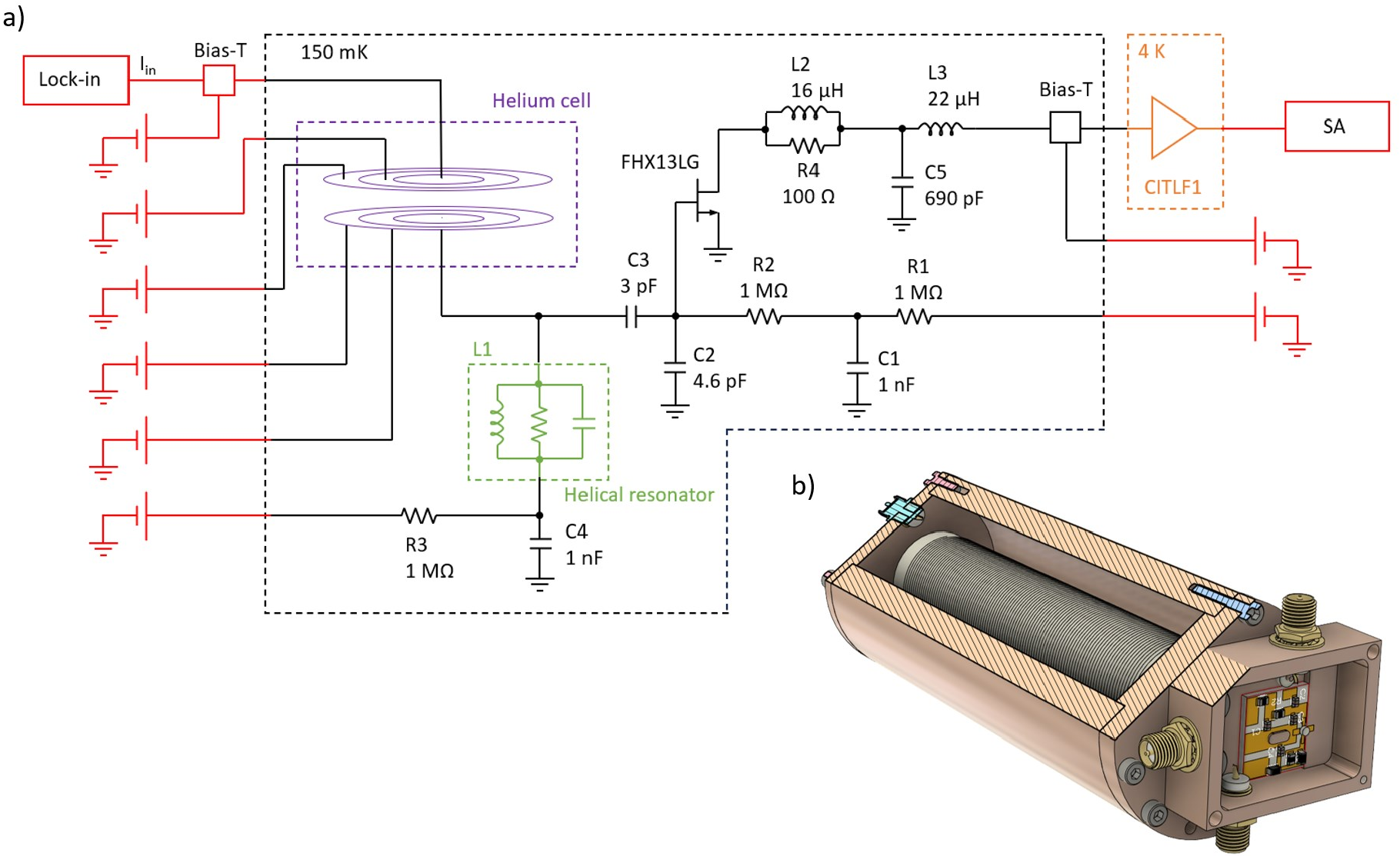}
	\caption{(color online)  (a) Principle schematics of the experimental setup and the amplifier circuit. (b) 3D view of the helical resonator (inductor L1) and FHX13LG HEMT amplifier.}
	\label{fig:3}	 
\end{figure}

The principal schematics of the experimental setup and the amplifier circuit are shown in Fig.~\ref{fig:3}(a). Our resonant amplifier consists of the LC tank circuit (the helical resonator) and FHX13LG HEMT amplifier, both located at the mixing chamber plate of a dilution refrigerators, followed by a commercial broadband cryogenic voltage amplifier located at the 4K plate~\cite{amp}. The tank circuit is connected to the detection electrode of the experimental setup described in details in the next section. Here, we provide description of the HEMT amplifier circuit. The gate bias voltage to HEMT is provided through two RC filters with 1 M${\rm \Omega}$ resistors. DC block capacitor C3 along with a capacitor C2 of second RC filter forms a voltage divider that adjusts the coupling of the transistor to the tank circuit comprised of an inductor L1. Resistor R3 and capacitor C4 form another RC filter for the DC bias of the detection electrode. A small inductor L2 in parallel with a resistor R4 inserts losses above the working bandwidth to suppress self-excitation of the amplifier. Finally, capacitor C5 and inductor L3 match output impedance of the transistor to a 50~${\rm \Omega}$ coaxial cable.  To exclude thermal drift of the amplifier parameters we used the thin-film resistors and NP0 ceramic multilayer capacitors. The amplifier components are assembled on a Rogers RO4000 Printed Circuit Board (PCB). The PCB is housed in a copper case with SMA connectors (see Fig.~\ref{fig:3}(b)) directly attached to the copper shielding of the inductor L1 to minimize the stray capacitance. To provide an efficient heat sink, one of the transistor source leads is soldered directly to the case. Additional heat sink is provided by copper mesa on the case bottom which goes through a hole in PCB and is soldered to the top of the PCB.

The inductor L1, which forms a helical resonator, is winded on a PTFE cylindrical core of 23 mm diameter and 100 mm length. To ensure the proper thermalization of the inductor we used a copper clad NbTi wire with formvar insulation. The bare diameter of the wire is about $115$~$\mu$m, with Cu to NbTi ratio 1.5:1. With 450 turns and with a 6.3 mm total length of winding we achieved an inductance and a series resistance of 924~${\rm \mu}$H and 93~${\rm \Omega}$, respectively, at the room temperature. The 3D view of the helical resonator inside the copper shielding is shown in Fig.~\ref{fig:3}(b). 

\subsection{Image charge detection of the Rydberg transition}
\label{sec:image}
        
To characterize the performance of our amplifier we have measured the image-charge signal induced by the Rydberg transition of surface electrons in response to a pulse-modulated microwave (MW) excitation. The experimental setup used for electrons on helium is similar to that described previously~\cite{KawaPRL2019}. A vacuum-tight cylindrical copper cell contains two circular conducting plates forming a parallel-plate capacitor with a 2 mm gap. Each plate consisted of three concentric electrodes separated by a narrow gap used to detect the presence of electrons by the conventional Sommer-Tanner method~\cite{SommPRL1971}. The cell is cooled down at the mixing chamber plate of the dilution refrigerator. Liquid helium is condensed into the cell and its level is set in the middle between the capacitor plates. Electrons are produced by the thermionic emission from a tungsten filament and are trapped on the liquid surface above the central (detection) electrode of the capacitor bottom plate by applying a positive DC bias voltage to the electrode. In the experiment described here all other electrodes were kept at zero potential. The transition between the ground state and the first-excited Rydberg state of the surface electrons are excited by applying the resonant MW radiation with the frequency $f$ that matches the transition frequency. When a pulse-modulated MW excitation at the modulation frequency $f_m$ is applied, an AC image current at the frequency $f_m$ is induced at either of the central electrodes of the capacitor plates by the electron transitions. As shown previously~\cite{KawaPRL2019}, this current can be directly measured with a lock-in amplifier by connecting its current input to the central electrode of the top capacitor plate (see Fig.~\ref{fig:3}(a)). Alternatively, the image current of the same magnitude and opposite sign can be measured using the resonant amplifier described in the previous section by connecting the amplifier input (the tank circuit) to the central electrode of the bottom capacitor plate. By measuring the input image current directly using the first method we can estimate the gain of our resonant amplifier by analyzing its output.  

\begin{figure}
	\includegraphics[width=1.0\textwidth]{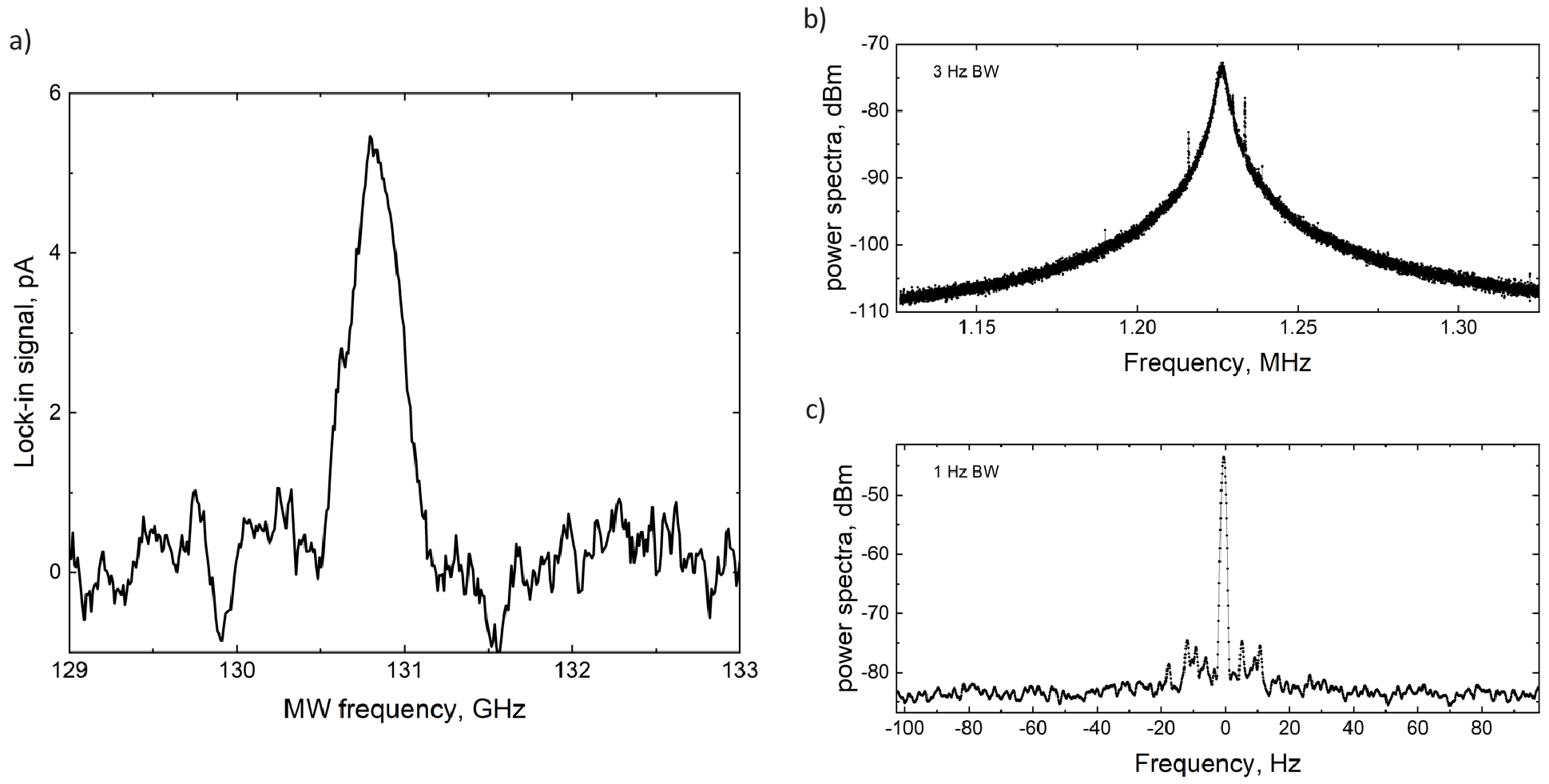}
	\caption{(color online)  (a) The Rydberg transition signal detected by measuring the induced image current directly at the central electrode of the capacitor top plate. (b) Noise spectrum of the resonant amplifier measured at the output of the 4K-stage broadband amplifier. (c) The Rydberg transition signal detected by the resonant amplifier on top of its noise spectrum. For clarity, the frequency on the horizontal axis is offset by the value of the tank-circuit resonance frequency $f_r=1.21898$~MHz.}
	\label{fig:4}	 
\end{figure}

The result of the measurements by the first method is shown in Fig.~\ref{fig:4}(a). Here, the image current induced at the top electrode is measured as a function of the MW frequency $f$ with about $5.5\times 10^6$ surface electrons confined between the capacitor plates. The measurements are done with the pulse-modulated MWs at the modulation frequency $f_m=1.4$~MHz. Note that the value of the modulation frequency was chosen to be somewhat detuned from the resonant frequency of the tank circuit connected to the bottom plate of the capacitor in order to exclude its excitation, which could affect the measurement results. At the Rydberg transition resonance corresponding to $f\approx 130.7$~GHz we estimate the image current due to the excited electrons of about $5.5$~pA. In the second method, the induced image current is measured at the bottom electrode while the output of the resonant amplifier is connected to the spectrum analyzer (SA), see Fig.~\ref{fig:3}(a). For this measurements, the HEMT amplifier was DC biased at $V_{gs}=-0.437$~V and $V_{ds}=0.25$~V ($I_d=31$~${\rm \mu}$A). For reference, Fig.~\ref{fig:4}(b) shows the noise spectrum of the resonant amplifier in the absence of the MW excitation. The maximum of the noise spectrum corresponds to the resonant frequency of the LC tank circuit $f_r=1.21898$~MHz with the circuit loaded quality factor $Q\approx 360$. To detect the Rydberg transition, the pulse-modulated MW excitation is applied at the resonant frequency $f=130.7$~GHz and the modulation frequency $f_m$ equal to the tank circuit resonace frequency $f_r$. Assuming that the image current is proportional to the modulation frequency $f_m$, we expect that the image current induced at the resonant amplifier is about $I_{i}=4.8$ pA.  At the tank-circuit resonance the circuit impedance is real and is given by $R_{LC}= 2\pi f_r QL\approx 2.55$~M${\rm \Omega}$. The corresponding voltage across the tank circuit due to the image current is $V_{LC} = I_{i} R_{LC}=12.2$~$\mu$V. As shown in Fig.~\ref{fig:2}(a), the voltage signal from the tank circuit is coupled to the input (gate) of HEMT through a voltage divider comprised of the capacitors C2 and C3 with the coupling ratio $\kappa = C_3/(C_2+C_3)=0.4$, thus giving the estimated voltage signal $\kappa V_{LC}= 4.8$~$\mu$V at the HEMT input. Fig.~\ref{fig:4}(c) shows the corresponding output signal after the 4K-stage voltage amplifier measured by the spectrum analyzer with 1~Hz resolution bandwidth (BW). On top of the noise spectrum we clearly observe a signal (with the signal-to-noise ratio ${\rm SNR}\approx 40$~dB) due to the image current with the frequency $f_m=f_r$ at the amplifier input. From the signal amplitude of about 1.5~mV$_{\rm rms}$ corresponding to the power amplitude of $-43.3$~dBm in Fig.~\ref{fig:4}(c) we estimate the voltage gain of our resonant amplifier of about $49.9$~dB, with the transimpedance gain of about 3.2~nA/V.  

\subsection{Noise and sensitivity estimations}
\label{sec:noise}

\begin{figure}
	\includegraphics[width=1.0\textwidth]{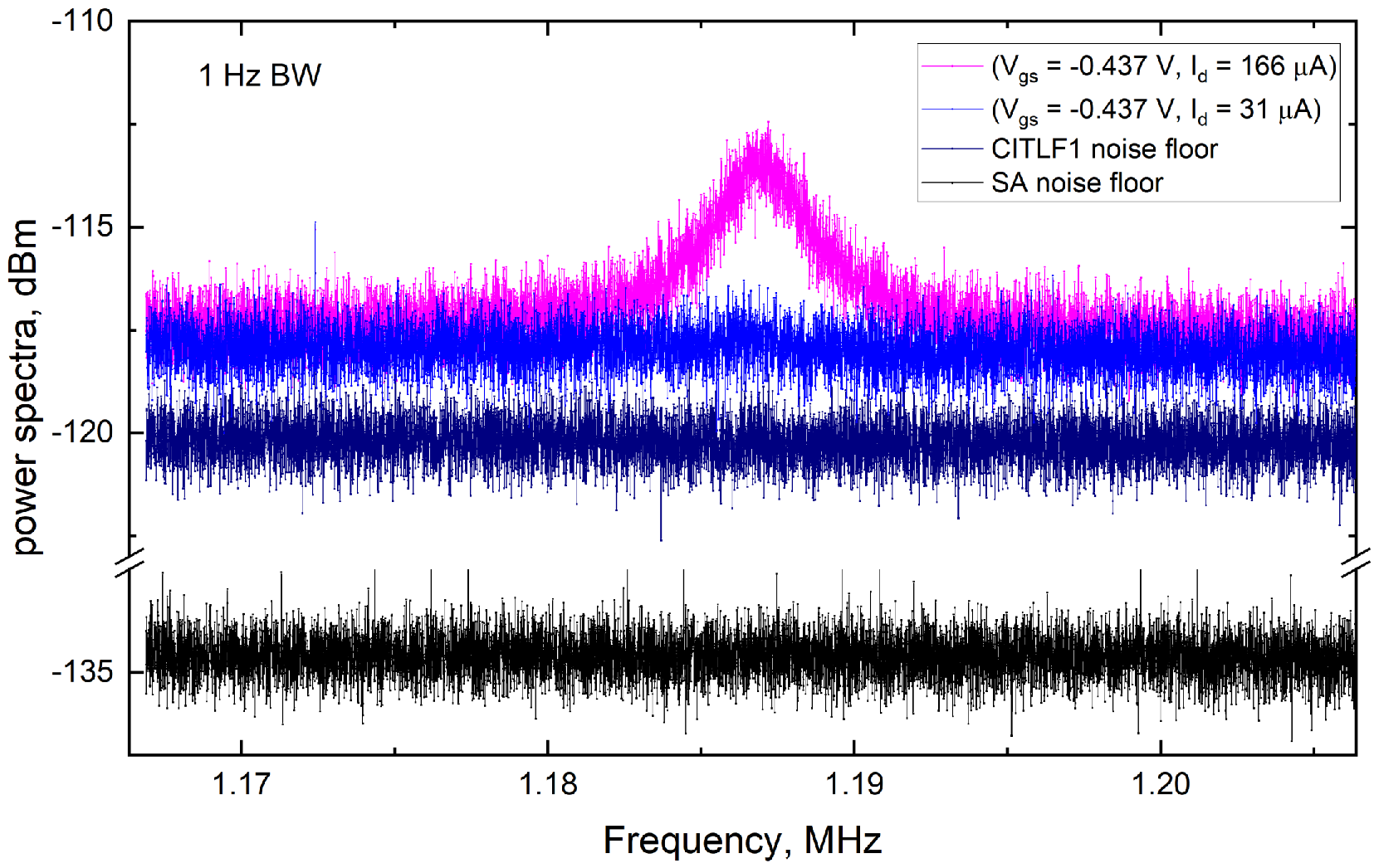}
	\caption{(color online)  Power spectra of the output of the 4K-stage amplifier measured by the spectrum analyzer (SA) within the 1 Hz resolution bandwidth (BW). Different traces are described in the text.}
	\label{fig:5}	 
\end{figure}
        
To characterize the sensitivity of our method towards detecting the Rydberg transition of a single electron we have to consider the sources of noise and possible ways of its reduction. The output noise of our resonant amplifier comes from the noise produced by the experimental setup at the amplifier input and the noise produced by each of its stages (the tank circuit, HEMT, and the broadband 4K-stage amplifier). By analyzing the output noise spectrum of the amplifier shown in Fig.~\ref{fig:4}(b) we concluded that it was dominated by the noise at the amplifier input from the experimental cell, such as the thermal noise leaking into the cell through the coaxial cables used for the DC biasing of the capacitor electrodes. In the subsequent experiments, it should be possible to eliminate or significantly reduce this noise by employing DC wiring and low-pass filtering of the DC lines. To characterize the noise of the amplifier itself we carried out an additional experiment with the amplifier disconnected from the cell. To obtain the same resonant frequency of the loaded tank circuit, we have added a 10 pF capacitor parallel to the inductor L1 to account for the stray capacitance of the disconnected cell. Fig.~\ref{fig:5} shows the noise spectra measured at the 4K-stage amplifier output by the spectrum analyzer with the 1 Hz resolution bandwidth. The lowest (black) trace shows the noise floor of the spectrum analyzer measured with both HEMT and the 4K-stage amplifier turned off. The second-lowest (dark blue) trace shows the noise spectrum with the 4K-stage amplifier turned on, while the HEMT amplifier is turned off. Finally, the two upper traces show the noise spectrum measured with the DC drain-bias $V_{ds}=0.25$~V, $I_d=31$~${\rm \mu}$A (blue trace) and $V_{ds}=0.5$~V, $I_d=166$~${\rm \mu}$A (magenta trace) applied to HEMT. To analyze these results we follow Ref.~\cite{UlmNuc2013} and represent the  noisy amplifier as an ideal amplifier with the voltage ($\bar{e}_n$) and current ($\bar{i}_n$) noise sources connected in series and in parallel, respectively, to the HEMT input. In general, the RMS (voltage) noise spectral density at the output $\bar{e}$ can be calculated as contribution from all amplifier stages, with the corresponding amplification coefficients, according to

\begin{equation}
\bar{e}^2 = (G_1 G_2)^2 \bar{e}_1^2 + G_2^2 \bar{e}_2^2 + \bar{e}_s^2, 
\label{eq:1}
\end{equation}

\noindent where $\bar{e}_1$ and $\bar{e}_2$ is the input noise spectral density at the HEMT and 4K-stage amplifier, respectively, $G_1$ and $G_2$ is the corresponding voltage gain of each amplifier, and $\bar{e}_s$ is the noise spectral density of the spectrum analyzer. Because the noise floor of the spectrum analyzer is at least $14.4$~dB (about 27.5 times in power) lower than the noise with the amplifiers turned on, the last term on the right-hand side of Eq.~(\ref{eq:1}) will be omitted from the further analysis. The second term on the right-hand side of Eq.~(\ref{eq:1}) can be determined from the power spectral density given (in dBm) by the second-lowest trace in Fig.~\ref{fig:5}, while the left-hand side of Eq.~(\ref{eq:1}) can be determined from the power spectral density given (in dBm) by the upper traces in Fig.~\ref{fig:5}. Considering the data in Fig.~\ref{fig:5} for the DC drain-bias $V_d=0.25$~V we obtain $G_2\bar{e}_2\approx 0.22$~${\rm \mu}$V and $\bar{e}\approx 0.28$~${\rm \mu}$V. By knowing the total voltage gain $G_1G_2\approx 312$ from the previous section, it is straightforward to calculate from Eq.~\ref{eq:1} the HEMT input voltage noise spectral density $\bar{e}_1\approx 0.6$~nV/$\sqrt{\rm Hz}$, with the assumed measurement bandwidth of 1 Hz. 

According to Ref.~\cite{UlmNuc2013}, the HEMT input noise $\bar{e}_1$ at the tank-circuit resonance is the combination of the voltage noise $\bar{e}_n$, the voltage noise across the tank circuit $\kappa^2\bar{i}_nR_{LC}$ due to the current noise $\bar{i}_n$, and the thermal noise of the tank circuit $\kappa\sqrt{4k_BR_{LC}T}$~\cite{UlmNuc2013}. For the DC drain-bias $V_{d}=0.25$~V (see Fig.~\ref{fig:5}) we did not observe any noticeable increase of the noise level at the tank-circuit resonance. This indicates that both the thermal noise of the tank circuit and the input current noise are smaller than the calculated voltage noise. That allows us to set a limit for the input current noise to be below 1.5 fA/$\sqrt{\rm Hz}$. However, note that the direct estimation of the thermal noise from the tank circuit at resonance is about $1.5$~nV/$\sqrt{\rm Hz}$, which is higher than the estimated $\bar{e}_1$. We believe this contradiction can be caused by using of oversimplified lump circuit model for power transfer from the experimental setup to the amplifier input. Contrarily, for the DC drain-bias $V_{d}=0.5$~V (the top trace in Fig.~\ref{fig:5}) we can see a noticeable increase of the noise level at the resonance. This indicates that either the thermal noise of the tank circuit or the current noise $i_n$ of the HEMT increases significantly with increasing $V_d$. This information is important for choosing the DC biasing of HEMT for the optimal sensitivity performance. 

The main goal of this study is to estimate feasibility of detecting the Rydberg transition of a single electron coupled to the tank circuit of our resonant amplifier. As shown recently, the image charge signal from the excited surface electrons can be significantly enhanced by confining them in a microchannel device, thus bringing them closer to the detection electrode~\cite{ZouNJP2022}. For a typical microchannel device employed in such experiments an electron is located at a distance $d\sim 1$~${\rm \mu}$m from the electrode, which gives an enhancement in the image-current signal by the factor $\sim 10^3$ comparing with the parallel-plate capacitor setup described in this work. For a single electron one estimates an image current $I_i=(\pi e \Delta z f_m)/D\approx 5$~fA due to the single-electron transition between the ground state and the first excited Rydberg state, which corresponds to $\Delta z=10$~nm, and $f_m = 1$~MHz. According to Sec.~\ref{sec:image} such current will produce an input voltage signal $\kappa I_iR_{LC}\approx 5$~nV. Comparing with the above estimate of the voltage noise we conclude that the single-electron Rydberg transition in a microchannel device can be detected with the signal-to-noise ratio SNR=8 and the measurement bandwidth 1 Hz, providing that the noise level is limited by the noise of the amplifier.   

\section{Conclusion}
\label{Conclusion}

In this work, the HEMT transistor FHX13LG was characterized down to 100 mK temperatures. By combining HEMT with a helical resonator, we have designed a narrow-band high-impedance resonant amplifier for the image charge detection. Based on the experimental observation of the image-current induced by the microwave-excited Rydberg transition in a many-electron system on liquid helium we determined the amplification factor and  estimated the noise of the amplifier. At the chosen DC biasing of HEMT transistor we determined the input voltage noise of 0.6 nV/$\sqrt{\rm Hz}$ and the input current noise below 1.5 fA/$\sqrt{\rm Hz}$. Based on these results we estimated the feasibility of detecting of single-electron Rydberg transitions in a 1${\rm \mu}$m-deep microchannel device and concluded that such a detection with SNR = 8 is achievable in the measurement bandwidth of 1 Hz, assuming a proper attenuation of the external noise. Further improvement of the amplifier sensitivity are feasible by increasing the quality factor of the helical resonator, e.g. by employing uncladded NbTi wire and Nb shielding to achieve the quality factor above $10^3$. 

\begin{acknowledgements}
The work was supported by an internal grant from the Okinawa Institute of Science and Technology (OIST) Graduate University and Grant-in-Aid for Scientific Research (Grant No. 23H01795) KAKENHI MEXT. J. A. K. acknowledges support of an OIST Graduate University internship program.
\end{acknowledgements}




\end{document}